\documentclass[pdflatex,sn-mathphys-num]{sn-jnl}%

\usepackage{graphicx}%
\usepackage{multirow}%
\usepackage{amsmath,amssymb,amsfonts}%
\usepackage{amsthm}%
\usepackage{mathrsfs}%
\usepackage[title]{appendix}%
\usepackage{xcolor}%
\usepackage{textcomp}%
\usepackage{manyfoot}%
\usepackage{booktabs}%
\usepackage{algorithm}%
\usepackage{algorithmicx}%
\usepackage{algpseudocode}%
\usepackage{listings}%
\theoremstyle{thmstyleone}%

\theoremstyle{thmstyletwo}%

\theoremstyle{thmstylethree}%

\begin{document}

\title[Article Title]{Infrastructure and its Discontents: Contestations of Power in Canada’s AI Governance Scaffold}

\author*[1]{\fnm{Christelle} \sur{Tessono}}\email{christelle.tessono@mail.utoronto.ca}

\author[1]{\fnm{Nusrat Jahan} \sur{Mim}}\email{nusrat.mim@daniels.utoronto.ca}

\author[1]{\fnm{Shion} \sur{Guha}}\email{shion.guha@utoronto.ca}

\affil*[1]{\orgname{University of Toronto}, \country{Canada}}

\abstract{This paper develops a critical theoretical framework for understanding national Artificial Intelligence (AI) strategies through the lens of ``Governance as Infrastructure." Integrating insights from Infrastructure Studies, Science and Technology Studies (STS), this paper employs Black Feminist Technoscience (BFTS) to reveal how AI governance infrastructures are assembled, maintained, and contested, and whose forms of knowledge, labour, and experience are rendered visible or invisible in these processes. We analyze the Canadian AI governance landscape through in-depth interviews with bureaucrats, civil society advocacy groups, academic researchers, and industry actors. Our inductive thematic analysis reveals how AI governance infrastructure enacts specific configurations of power, particularly through the politics of expertise; how actors navigate profound ambiguities related to AI's opacity, often involving strategic ignorance; and how mechanisms for justice and accountability struggle to address AI-related harms. We contend that a BFTS perspective highlights the limitations of current governance approaches and points toward alternative infrastructural designs prioritizing care, maintenance, participatory democracy, and adaptive temporalities to foster more equitable and accountable technology governance.}


\maketitle

\section{Introduction}\label{sec1}\label{sec1}
In 2019, the Canadian Food Inspection Agency (CFIA) adopted an AI-driven risk-ranking system to prioritize food safety inspections. Yet by late 2024, a deadly listeria outbreak went largely undetected due to insufficient testing and under-reporting~\cite{robertson_algorithm_2024}. Meanwhile, six Black refugees—five of them women—had their status revoked by Immigration, Refugees, and Citizenship Canada (IRCC)\footnote{IRCC is a federal department which manages refugee protection, immigration, and citizenship services in Canada.} after a facial recognition system misidentified them without their knowledge, forcing them to appeal their cases to the Federal Court~\cite{christian_how_2023}. Despite Canada’s global AI leadership through institutes like Vector, Mila, and Amii, and ongoing collaborations with academia and industry, the newly elected Prime Minister, Mark Carney, plans to scale AI use in public services without a uniform regulatory framework in place~\cite{carney_mandate_2025}. An increasing pattern of harmful incidents, alongside efforts to deploy AI in the absence of a regulatory framework, has prompted civil society actors to raise serious concerns about the government's strategy~\cite{british_columbia_civil_liberties_association_open_2025}. These incidents reveal cracks in Canada’s AI governance infrastructure, from its oversight gaps to the deployment of harmful biased systems. Combined with accelerated AI adoption across public and private sectors, these fissures highlight the urgent need for a new framework to understand and address such failures.

To explore Canada’s AI governance challenges, we develop a novel critical framework of ``AI Governance as Infrastructure," integrating Infrastructure Studies, Science and Technology Studies (STS), and Black Feminist Technoscience (BFTS). This framing goes beyond treating policy as a set of abstract rules and practices, and instead sheds light on the hidden systems, actors, routines, and power relations that undergird AI governance. By doing so, our study fills a gap in current AI governance research, which has rarely been examined through an infrastructural and Black Feminist lens. Moreover, while critical scholars within STS, related feminist and constructivist approaches have productively examined how technologies operate and are regulated, they often privilege moments of design, innovation, or formal decision-making as sites of inquiry~\cite{suchman_located_2002, wajcman_feminist_2010, hans_k_klein_social_2022}. Our study builds from this by shifting analytical attention away from moments of technological development and adoption toward the everyday work of maintaining, contesting, and repairing AI governance infrastructures, thereby revealing forms of power, harm, and responsibility that remain largely invisible within existing critical approaches.

This paper uses Canada as a case study by drawing on in-depth interviews conducted with the country's technology policy community, from researchers in academia, public servants in federal and provincial government departments and regulatory offices, civil society organizations and industry actors. We examine the challenges that render governing generative AI systems in Canada ineffective, inequitable and unaccountable. Our study seeks to address the following research questions: 

\begin{enumerate}
    \item \textbf{RQ1} How does Canada's AI governance infrastructure enact power?
    \item \textbf{RQ2} How do rapid advancements in AI challenge Canada's AI governance infrastructure? In particular, how do ambiguities created by AI's opacity complicate governance responses?
    \item \textbf{RQ3} How does Canada's AI governance infrastructure address (or fail to address) these challenges, given AI-specific harms like scaled discrimination and environmental degradation?
\end{enumerate}

This paper begins in Section~\ref{sec2} by outlining a working definition of AI, followed by a brief review of the literature on AI governance and Infrastructure Studies. In Section~\ref{secBFTS}, we theorize Black Feminist Technoscience (BFTS) and provide a critical framework for assessing ``AI Governance as Infrastructure". In Section~\ref{sec3}, we then provide a brief overview of Canada’s current approach to AI. In Section~\ref{sec4}, we outline our study's methodology. In Section~\ref{sec5}, we present our empirical findings. In Section~\ref{sec6}, we apply our critical assessment framework to this study's findings. And lastly, in Section~\ref{sec7}, we outline resulting policy recommendations that can be applied in the Canadian at various levels of governance, as well as in other domestic contexts. The paper concludes by reflecting on its limitations and identifying directions for future research. While focused on the Canadian context, this analysis offers insights that may be relevant to AI governance efforts in other jurisdictions. Our study reveals how AI governance infrastructures reproduce power dynamics via expert politics, navigate AI’s opacity, struggle with justice and accountability, and are shaped by competing visions of what AI should achieve and whose values it should serve. The BFTS lens exposes current shortcomings and suggests alternative governance designs prioritizing care, participatory democracy, and adaptive timelines for equitable and accountable tech futures.

\section{Related Work}\label{sec2}\label{sec2}

\subsection{Defining AI}\label{defining_ai}
We recognize that AI is a term which has been ``increasingly saddled with hype, misdirection, and confusion”~\citep[256]{veale_ai_2023}. As such, for the purposes of this study, we adopt a definition of AI that is broad and inclusive of the actors, processes, and resources necessary to develop and maintain these systems. Particularly, we draw on Crawford (2021) who defines AI as being ``both embodied and material, made from natural resources, fuel, human labor, infrastructures, logistics, histories, and classifications” and ``depends entirely on a much wider set of political and social structures” (8). We also draw on Attard-Frost's (2023) interpretative frames, which note that an ecosystem view defines AI as ``as an interdependent set of computational, social, technical, and material components that integrate across multiple scales and contexts” (15). As such, we view AI as a sociotechnical system materialized through the economic, political and social arrangements between networks of labor (e.g. software engineers, custodians, data workers, legal counsel, etc.), natural (e.g. water, fossil fuels, metals, land, etc.), material (e.g. fiber optic,  etc.), computational (e.g. algorithms, data, graphics processing units, etc.), and physical resources (e.g. data center facilities, content moderator offices, etc.). 

This definition is intentionally broad in order to not solely focus on the computational materials, tools, and techniques necessary for an AI system's operations. AI systems operate at a scale that distinguishes them from other information technologies in terms of data volume, infrastructure and resource demands. In other words, this definition includes, but is not limited to machine learning and instead encompasses the application of these computational techniques in systems such as facial recognition technology, disease diagnosis systems, risk prediction software, livestock health monitoring, drone-assisted surveillance, voice assistants, automated translation tools, and chatbots to name a few examples. Our intention is not to treat these applications as equivalent, but to instead recognize the inherent network of social, legal, material, and environmental resources and relations necessary to the development and maintenance of AI systems that they share in common.

And lastly, we chose to adopt this definition as it allows us to include the different types of harms that may occur along the lifecycle of AI systems, as well as the diversity of actors involved in building, maintaining, procuring and using these technologies (e.g. data workers, regulatory compliance officers, counsel, public servants, students, doctors, custodial staff, teachers, etc.). These harms may include, but not be limited to, environmental harms (e.g. data center clean water consumption), labor harms (e.g. wage theft, data worker exploitation, etc.), and informational harms (e.g. spread of misinformation on social media platforms) to name a few examples.

\subsection{Governance As Infrastructure}\label{subsec1}
We now examine the body of literature at the intersections of technology governance and infrastructure. Rather than treating them as abstract domains or collections of policies, we approach them through the lens of infrastructural theory. Infrastructure has long held a central place in Science and Technology Studies (STS), where it is conceptualized not merely as a physical or technical foundation, but as a sociotechnical assemblage that sustains and organizes everyday life~\cite{larkin_politics_2013}. Scholars have argued that infrastructure is not a thing, but a relationship: it becomes infrastructure only in relation to organized practices and when it supports activities across time and space~\citep{star1994steps, star_steps_1996}. These systems often remain invisible until they break down, at which point their social and political dimensions come sharply into focus~\citep{star_ethnography_1999}. Infrastructure, in this view, is not static. It is constantly shaped by routines, standards, classifications, and the work of alignment among diverse actors~\cite{stanforth2006using, howe2016paradoxical}.

This understanding allows scholars to analyze infrastructures not just as neutral backdrops, but as deeply political and epistemic formations. Bowker and Star (1999) highlight how infrastructures encode assumptions about whose knowledge counts and how categories of people and things are made actionable~\citep{bowker_sorting_1999}. Similarly, Singh and Jackson (2021) introduce the idea of “seeing like an infrastructure” to describe how bureaucratic systems and data schemas function as tools of both legibility and exclusion, enabling state vision while simultaneously shaping how citizens appear, participate, or remain invisible~\citep{singh2021seeing}. These works collectively foreground the often hidden labour of infrastructural maintenance, as well as the normative struggles embedded in classification systems, technical standards, and institutional workflows.

Building on the relational view of infrastructure, scholars have increasingly conceptualized governance itself as a form of infrastructure, a distributed system of norms, routines, institutions, and tools through which social order is produced and maintained~\cite{mcfarlane2008political, hinings2017fields, pham2023policy}. In this framing, governance is not merely about rule-making or enforcement; it is about the embedding of authority, responsibility, and decision-making into durable, routinized systems that shape everyday life. These governance infrastructures consist of institutional protocols, regulatory procedures, and organizational workflows that mediate power and shape the very conditions under which political and technical decisions are made~\citep{star_steps_1996, bowker_sorting_1999}. By treating governance as infrastructure, we can trace how authority is operationalized not just through formal statutes but through mundane practices, such as standards, reporting tools, classifications, licensing procedures, that enable or constrain action~\cite{hammerschmid2017governance, winner2017artifacts, lipsky2010street}.

Viewing governance as infrastructure offers key analytic benefits. It reveals how power operates not just in formal politics but through routine, often invisible systems~\cite{star1996steps}, highlights governance as a temporal process of institutional maintenance and adaptation~\cite{ahmed2015residual}, and exposes how stabilized mechanisms become politically invisible and resistant to critique~\citep{star_ethnography_1999}. This perspective also draws attention to what Bowker and Star (1999) call ``torque", the friction that arises when institutional categories misaligned with lived experience or when maintenance favours stability over equity~\citep[27]{bowker_sorting_1999}.

\subsection{Seeing AI Governance as Infrastructure}\label{subsec2}
Viewing governance as infrastructure becomes especially urgent in the context of AI, where the pace of technological innovation consistently outstrips the capacity of institutions to respond. AI governance can be broadly defined as the evolving array of rules, norms, tools, and institutional mechanisms designed to align AI development and deployment with public values and collective goals~\cite{mantymaki_defining_2022, attard-frost_ai_2024, black_regulating_2019}. From international agreements to municipal legislation, corporate codes of conduct, and voluntary guidelines, AI governance has proliferated across jurisdictions and sectors~\cite{arnold_introducing_2024,kasirzadeh_algorithmic_2022}. These efforts are often driven by a growing recognition of the adverse impacts of AI systems, including discrimination, environmental degradation, labour precarity, and privacy violations, which have prompted the emergence of ethical charters, principles, and technical safeguards~\cite{koniakou_rush_2023, vesnic-alujevic_societal_2020}. Yet these normative efforts, while rhetorically powerful, often lack specificity, enforceability, and operational clarity~\cite{koniakou_rush_2023, vesnic-alujevic_societal_2020}. Ethical frameworks tend to condense complex sociotechnical issues into broadly appealing themes, making them communicable but ultimately inadequate for regulating AI practice~\cite{koniakou_rush_2023}. 

To move beyond this impasse, scholars have called for more comprehensive regulatory regimes that integrate ethical norms with enforceable legal structures, clearly delineated responsibilities, and diverse regulatory actors~\cite{black_regulating_2019, maclure_ai_2025}. Such regimes often involve multiple forms of intervention: licensing, auditing, sanctions, impact assessments, standardization, and litigation \cite{attard-frost_ai_2024}. Governments deploy an array of professionals (e.g. analysts, legal counsel, regulators, technologists, tribunals, and auditors) each empowered to carry out investigative, corrective, and preventive functions across different stages of the AI lifecycle~\cite{black_regulating_2019}. These actors form the working parts of an emergent governance infrastructure, an assemblage of policies, institutions, and routines that operates unevenly, adapts incrementally, and is often shaped by negotiation and contestation.

However, the infrastructural conditions under which AI governance unfolds remain under-theorized — a gap our study seeks to address. The growing literature on AI policy and regulation has focused on defining principles or evaluating legal instruments, less attention has been paid to the procedural, material, and relational dynamics that enable (or hinder) governance in practice. These include challenges of bureaucratic capacity, fragmented accountability, institutional opacity, and the concentration of influence among powerful technology firms~\cite{taeihagh_assessing_2021, taeihagh_governance_2021}. The opacity of AI systems compounds these issues by making harms difficult to trace and regulate, while the asymmetrical power of major tech companies, often the very entities tasked with self-regulation, limits the effectiveness of oversight mechanisms~\cite{taeihagh_assessing_2021}. Moreover, the persistent reliance on ethical codes without binding enforcement tools has resulted in a governance model that privileges symbolic compliance over structural change~\cite{koniakou_rush_2023}. As such, viewing Canada’s approach through the lens of ``Governance as Infrastructure", we uncover how mundane policy tools and routines (e.g., funding programs, standards, etc.) encode power and values in practice.

While the infrastructural lens offers a powerful framework for analyzing how governance systems are materially enacted and institutionally stabilized, it is not without limitations. Scholars have argued that infrastructure studies often focus on embeddedness and routine function, which can unintentionally normalize the very systems they aim to critique~\citep{star_ethnography_1999, singh2021seeing}. Emphasizing infrastructural continuity and maintenance may obscure the fractures, refusals, and alternative imaginaries that challenge dominant modes of governance~\cite{graham2002splintering}. In particular, this approach tends to foreground durability and alignment while neglecting the politics of exclusion, the labour of survival, and the epistemic disqualifications that structure sociotechnical life~\cite{easterling2014extrastatecraft}. When applied uncritically, the infrastructural perspective can flatten structural power asymmetries and fail to account for how governance systems are often historically contingent~\cite{easterling2014extrastatecraft}. These omissions are especially problematic in the context of AI governance, where power is often distributed unequally~\cite{gasser2017layered,attardfrost2025ai}, and where policy and technical systems often reproduce long-standing systems of domination~\cite{ziosi2025rethinking} under the guise of neutrality or innovation~\cite{dominguez2025reclaiming}. 

Scholars across STS, Media and Communication Studies, and Black Studies have all made calls to critically engage with how systemic forms of oppression coalesce with information technologies\cite{fuchs_internet_2008, j_khadijah_abdurahman_technologies_2021, noble_algorithms_2018, benjamin_catching_2016, vaidhyanathan_afterword_2006, steve_breyman_sts_2017}. For instance, Laura Mamo and Jennifer Fishman (2013) urge STS scholars to extend their analyses beyond surface level ethical considerations, and instead focus on developing a more robust and explicit engagement with justice efforts through their research~\cite{mamo_why_2013, benjamin_catching_2016}. Collectively, these calls to action demand a turn toward theoretical frameworks that center justice and critically engage with the development and maintenance of information technologies such as AI, prompting our turn to a Black Feminist Technoscience perspective that moves beyond the limits of prevailing STS approaches.

\section{Theorizing A Black Feminist View of AI Governance}\label{secBFTS}
\begin{figure} 
    \includegraphics[width=\linewidth]{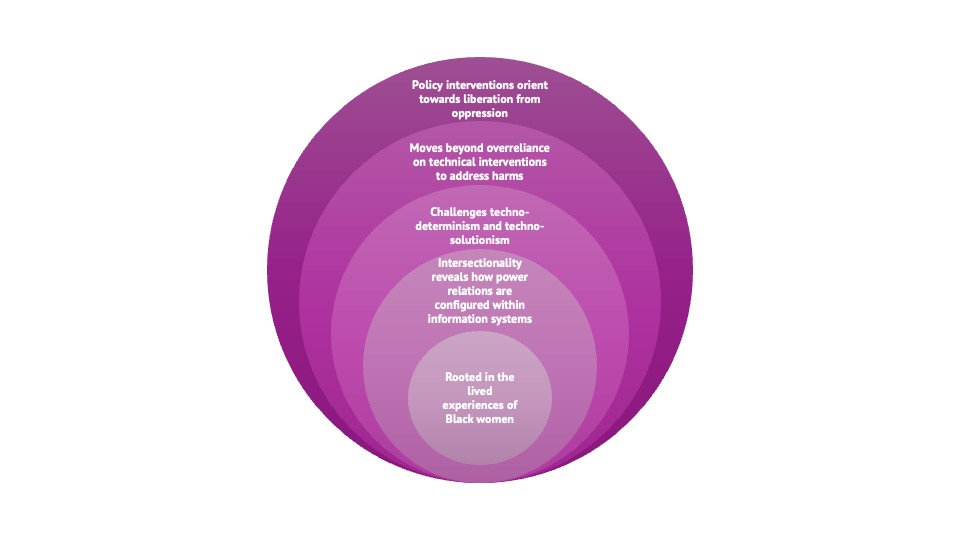}
    \caption{Conceptual dimensions of Black Feminist Technoscience informing the analytical framework used in this study.}
    \label{fig:bftdimensions}
\end{figure}

To address the limitations of infrastructure theory (and STS more broadly) in analyzing AI governance, we turn to Black Feminist Technoscience (BFTS) as a necessary and transformative framework. We build on Sherman's (2023) theorization of BFTS, which brings Black feminist and feminist technoscience scholarship in conversation to center Black women's knowledge and practices - thus challenging hegemonic assumptions about design and technology~\cite{sherman_black_2023}. Building on these contributions, we define BFTS as a meta-theoretical approach that interrogates the dominant frameworks that shape knowledge production, science, and technological development by centering Black Feminist Thought. Here, we adopt Harraway's definition of technoscience, which refers to the “dense nodes of human and nonhuman actors that are brought into alliance by the material, social, and semiotic technologies through which what will count as nature and as matters of fact get constituted for—and by—many millions of people”~\citep[50]{haraway_modest_1997} . Or in Wajcman's words, we use the term technoscience to denote how “technologies and new forms of social life are co-produced”~\citep[39]{wajcman_technofeminism_2004}. As depicted in Figure~\ref{fig:bftdimensions}, BFTS is rooted in the experiences of Black women, draws on intersectionality and the matrix of domination as core analytical tools. While intersectionality examines how systems of oppression intersect, the matrix of domination reveals how power manifests itself across these systems in people’s lives~\cite{hill_collins_black_2000}. In their literature review of STS journals, Grzanka and Bhatia (2023) find that intersectionality remains largely absent in mainstream STS, noting that its marginal presence significantly hinders critical scholarship~\cite{grzanka_intersectionality_2023}. Moreover, the authors note that while concepts like co-production and co-constitution acknowledge sociotechnical interactions, they lack an inherent focus on justice, a crucial aspect that intersectionality provides. Noble (2016) further argues that current STS frameworks often overlook the racialized and gendered dimensions of technology, limiting their ability to address the impact on marginalized groups, particularly on Black people~\cite{noble_future_2016}. These gaps and limitations highlight the need for the adoption of a framework grounded in BFTS to fully understand and intervene in systems of oppression.

When applying BFTS, scholars have been able to look at the role of colonialism and racial capitalism in shaping the conditions under which technology is developed and subsequently experienced \cite{hampton_techno-racial_2023}. For instance, this may look like shedding light on the hazardous working conditions of social media content moderators in Kenya~\cite{gebrekidan_content_2024}. When it comes to AI, BFTS can be adopted to understand how the design, development, deployment, and use of these technologies enact surveillance logics, carcerality, harmful and discriminatory search results to reinforce anti-Blackness~\cite{browne_dark_2015, benjamin_catching_2016, andrews_welfaire_2026, noble_algorithms_2018}. BFTS offers methodologies, theories, and pathways towards collective liberation which are grounded in decolonization, justice, care, accountability, and relational ethics, which in turn offer alternatives to hegemonic approaches to design and technology governance~\cite{birhane_algorithmic_2021, olojo_data_2026, harrington_all_2022}. In other words, BFTS uses the lived experiences of Black women, their ``everyday, taken-for-granted knowledge", to develop new meanings, raise consciousness, and build alternative worlds~\citetext{\citep[32]{hill_collins_black_2000},\citealp{lorde_uses_1997}}.

Although infrastructural perspectives help surface the material, procedural, and institutional layers of governance, they often fall short in capturing the structural exclusions~\cite{noble_algorithms_2018, jacobs1961death}, historical violences, and epistemic erasures~\cite{scott2020seeing,foucault2023discipline} that shape how governance systems are built and experienced. BFTS offers a critical expansion of this lens by foregrounding how race, gender, class, colonialism, and global capitalism influence the design, development, deployment, and regulation of technologies~\citep{noble_future_2016, benjamin_race_2019, floegel_methods_2022}. These intersecting forces are not peripheral to AI governance, they are foundational. BFTS challenges calls attention to the political economies, racialized labor, and extractive infrastructures that underpin how technologies, and the systems governing them, are imagined, assembled, and used~\cite{noble_future_2016}. As Benjamin (2019) notes, these design decisions, whether in technical architecture or policy formulation, are deeply political, shaped by libertarian values that privilege corporate autonomy and individualism over collective accountability~\cite{benjamin_race_2019}. These systems do not just mirror bias; they hide, accelerate, and deepen it, which in turn, contributes to a kind of ``digital denial”, a refusal to confront the structural origins of harm embedded in AI systems and their regulatory regimes~\citep[8-12]{benjamin_race_2019}.

\begin{figure} 
    \includegraphics[width=\linewidth]{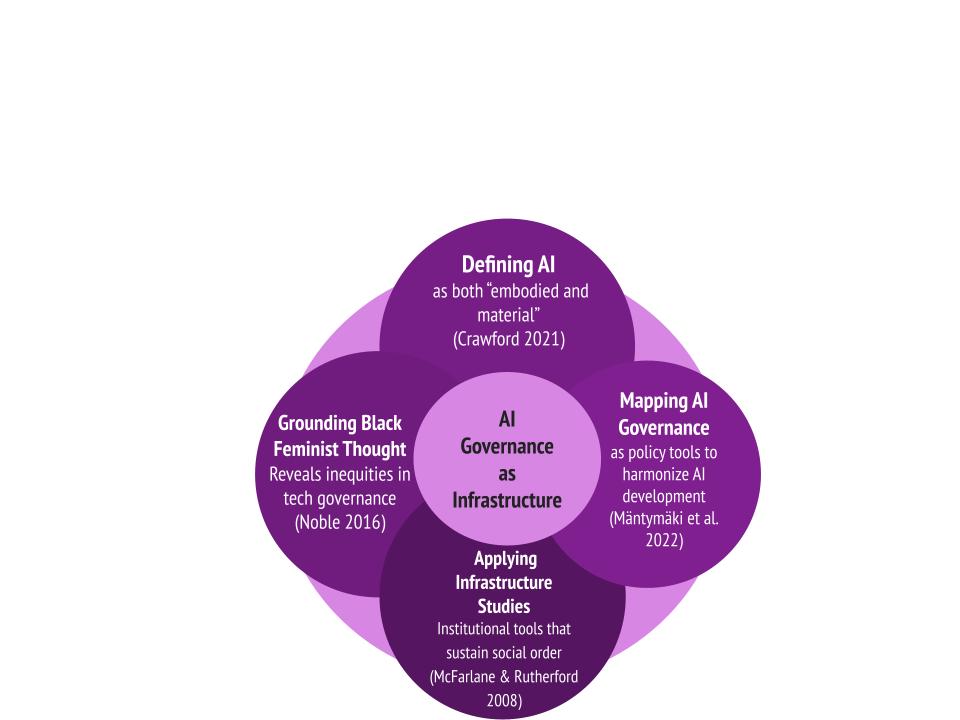}
    \caption{Overview of the theoretical framework of ``AI Governance as Infrastructure" developed to inform this research project.}
    \label{fig:cycle}
\end{figure}

This silence extends across much of the emerging AI governance literature. While scholars have offered important critiques of regulatory gaps, implementation failures, and ethical vagueness~\cite{arnold2024introducing, carpenter2024fda, bommasani2024foundation}, few have asked how governance itself might operate as a mechanism of exclusion, not just oversight~\cite{attardfrost2025ai}. As illustrated in Figure~\ref{fig:cycle}, BFTS fills this gap by offering a structural critique of power, showing how institutional decisions and epistemic frameworks determine who is rendered visible, governable, or expendable within AI systems. Beyond critique, BFTS enables a re-imagining of governance as a practice grounded in care and maintenance, and adaptive temporality. By putting BFTS in conversation with critical STS scholarship, we understand care as the relational and political labour that responds to systemic neglect~\cite{Stephen2026Care, finch_introduction_2022, puig_de_la_bellacasa_matters_2017, martin_politics_2015}; maintenance is the often invisibilized work of sustaining and re-imagining life amid failing systems\cite{kaba_we_2021, sharpe_wake_2016, star_ethnography_1999, gillespie_rethinking_2014}; and adaptive temporality is an ethical stance that resists the rush of innovation and makes space for repair and refusal~\cite{finch_introduction_2022, benjamin_viral_2022, benjamin_race_2019, sharpe_wake_2016}. Together, these commitments challenge dominant models that prioritize innovation and efficiency over accountability, offering a justice-oriented framework for re-imagining AI governance.

\subsection{Advancing A Critical Assessment Framework for AI Governance Infrastructures}\label{framework1}
By bringing BFTS into dialogue with infrastructure theory, this paper contributes a new lens to AI governance scholarship: one that centers care, maintenance, and adaptive temporality as essential to how governance systems are built, sustained, and experienced. Existing literature often overlooks these relational and reparative dimensions, focusing instead on formal rules or technical fixes \cite{benjamin_captivating_2019}. Our approach fills this gap by showing that effective and just AI governance requires not only regulation, but the everyday labor of repairing harm, sustaining trust, and attending to those most affected. Thus, our theoretical framework combines together infrastructure theory with BFTS to guide our analysis by revealing whose knowledge is privileged in AI policy, which issues are ignored or obscured, and where opportunities for care, refusal, and justice are being missed in Canada’s AI governance.

As discussed in Section~\ref{defining_ai}, we consider AI systems as being made up of not only technical components such as data, algorithms and computing infrastructure, but also of the actors and conditions involved in shaping them: from data workers, product managers, regulatory compliance officers, policymakers crafting standards, laws, guidelines, and codes. Building from Infrastructure Studies, we contend that these systems involve significant negotiation between actors, these negotiations are continuous flows that reflect how political, financial, ethical and social choices are prioritized by governments, industry and civil society~\cite{pickren_global_2018, star_how_2010, collier_post-soviet_2011}. Our literature review demonstrates that applying theories of Infrastructure Studies and Black Feminist Technoscience to AI governance allows us to understand the unique dynamics that shape policy debates and outcomes. By combining these theoretical perspectives, we advance a critical framework for understanding ``AI Governance as Infrastructure", which we apply to interpret and evaluate the insights from the interviews in Section~\ref{sec6}.

\begin{figure} 
    \includegraphics[width=\linewidth]{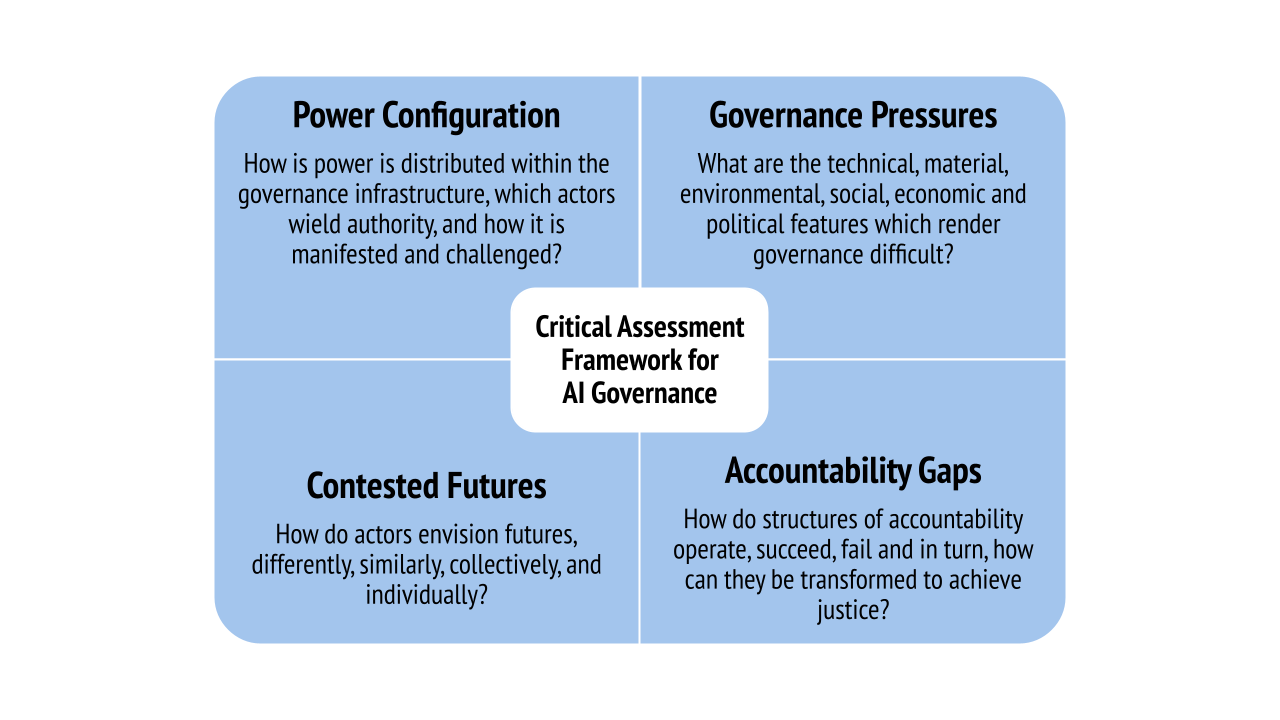}
    \caption{Matrix outlining the critical assessment framework developed for this project.}
    \label{fig:matrix}
\end{figure}

This framework's central claim is that when adopting the theoretical approach of ``Governance as Infrastructure" to AI, we are able to understand AI governance as a highly contested infrastructure, where power, knowledge, and justice are continuously negotiated. In turn, we adopt Black Feminist Technoscience to address these tensions by emphasizing care, refusal, maintenance, and adaptive temporality in policy configurations. As illustrated in Figure~\ref{fig:matrix}, the key dimensions of this critical assessment framework are as followed:
\begin{enumerate}
    \item \textit{Power Configuration}: Drawing from Collins' (2000) matrix of domination and Star \& Ruhleder (1994) conceptualization of the relational quality of infrastructure, this dimension examines how power is distributed within the governance infrastructure, which actors wield authority, and how it is manifested and challenged~\cite{hill_collins_black_2000, star1994steps}.
    \item \textit{Governance Pressures}: Guided by Pasquale's (2015) and Benjamin's (2019) discussions on opacity, as well as Bowker \& Star's (1999) analysis of how categories encode power, this dimension examines the technical, material, environmental, social, economic and political features which render governance uniquely difficult~\cite{pasquale_black_2015, benjamin_race_2019, bowker_sorting_1999}. 
    \item \textit{Accountability Gaps}: Incorporating Suchman's (2002) work on accountability, as well as Noble's (2016) analysis of the global chains of exploitation on which AI development relies on, this dimension examines how structures of accountability operate, succeed, fail, and in turn, how they can be transformed to achieve justice~\cite{suchman_located_2002, noble_future_2016}. 
    \item \textit{Contested Futures}: Building on Jasanoff \& Kim's sociotechnical imaginaries and Benjamin (2019) work on alternative technological futures, this dimension examines how actors envision futures, differently, similarly, collectively, and individually~\cite{jasanoff_dreamscapes_2015, benjamin_captivating_2019}. 
\end{enumerate} 

\section{Contextualizing Canadian AI Governance}\label{sec3}\label{sec3}
The Canadian approach to AI governance can be divided into two broad categories: i) federal investments and, ii) decentralized policy interventions. Federal investments include the Pan-Canadian AI Strategy, focusing on commercialization, standardization, and talent/research~\cite{1noauthor_pan-canadian_2024}. This includes \$125M (2021–2026) for AI research institutes, \$8.6M for AI standards, and \$208M over 10 years to attract researchers via the Canadian Institute for Advanced Research. The 2024 budget added \$2.4B for the Canadian AI Sovereign Compute Strategy, funding startups, small business AI integration, worker reskilling, and the Canadian AI Safety Institute~\cite{noauthor_securing_2024}. However, these investments face criticism for favouring corporate interests and regulatory capture, exemplified by Element AI’s public funding yet foreign acquisition ~\cite{brandusescu_artificial_2021, roberge_narvals_2022, bannerman_tech_2024}.

Policy-wise, Canada lacks a unified AI law, relying instead on fragmented rules for public and private sectors. Public sector AI is overseen by the Treasury Board Secretariat (TBS), which issues initiatives such as the AI Strategy for the Federal Public Service 2025-2027, the Directive on Automated Decision-Making (DADM), Algorithmic Impact Assessment (AIA), and the non-binding Guide on Generative AI~\cite{1secretariat_treasury_2015, 1secretariat_directive_2024, 1secretariat_guide_2024}. While these initiatives may point to a diverse set of efforts, critics highlight exclusions of national security, lack of recourse for harms, and poor transparency as key limitations for their effectiveness~\cite{scassa_administrative_2021, brandusescu_design_2025, karanicolas_err_2019, tag_starling1, das_bureaucratic_2026, das_is_2026}.. For instance, a study found that only 4\% of federal AI systems had publicly available impact assessments~\cite{tag_starling1}.  

In the private sector, no enforceable AI law exists. The Artificial Intelligence and Data Act (AIDA) under Bill C-27, introduced in 2022, lapsed due to prorogation and elections~\cite{canada_digital_charter}.\footnote{Prorogation ends a parliamentary session, halting all pending bills.} AIDA faced backlash for its minimal safeguards and lack of consultation~\cite{1csilc_key_2024, tessono_ai_2022, benson_devils_2023}. Additionally,  while there is a Voluntary Code of Conduct for Generative AI, it lacks enforcement~\cite{1voluntary_code_2024}. In sum, Canada’s federal AI governance prioritizes investment but falls short on enforceable regulation, leading to ineffective oversight. This context motivates our research questions: we investigate how such an AI governance infrastructure distributes power, copes with opacity, and addresses (or fails to address) AI-specific harms.

\section{Methods}\label{sec4}\label{sec4}\label{sec4}
\begin{figure}
    \centering
    \includegraphics[width=\linewidth]{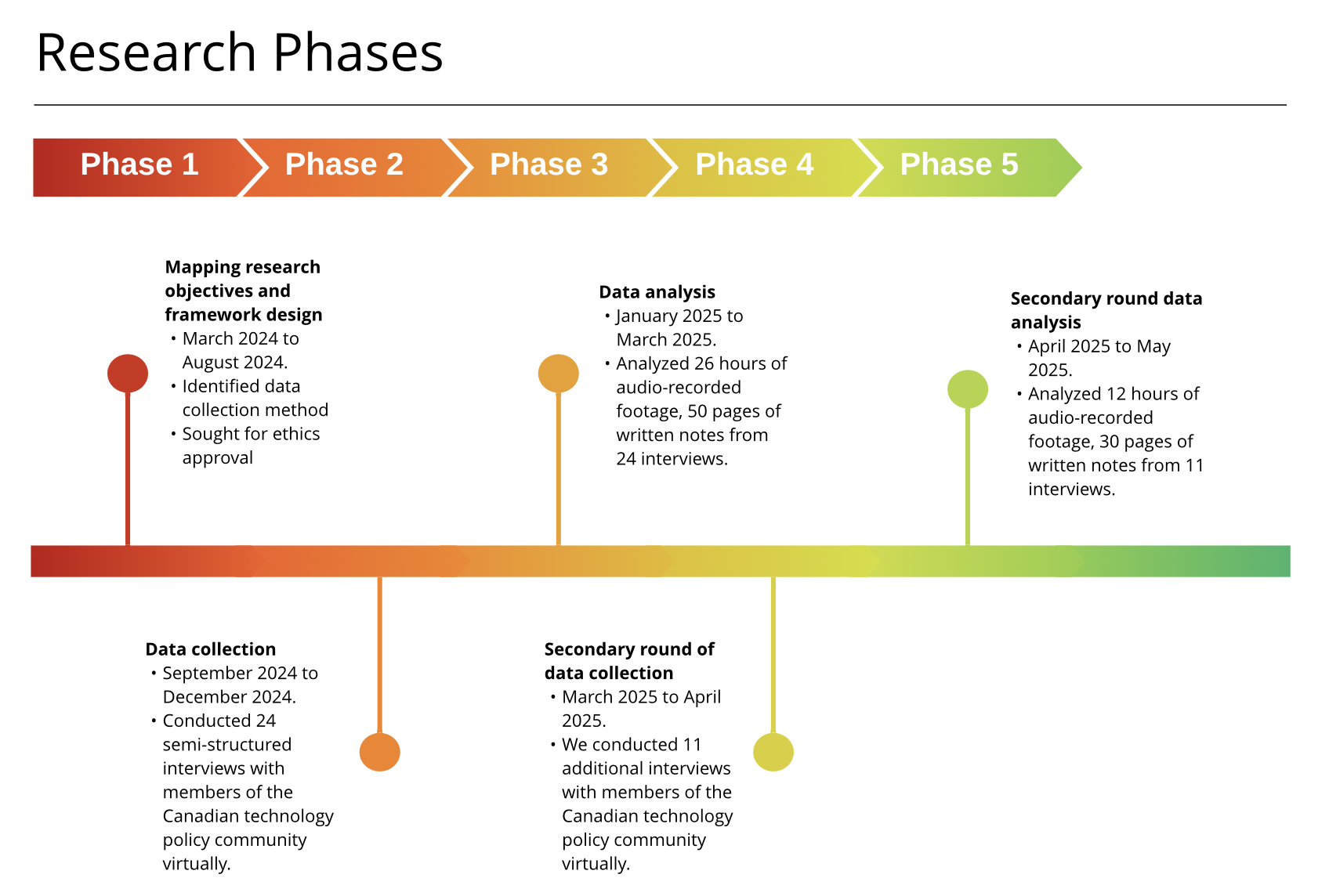}
    \caption{Timeline depicting the research phases and activities conducted over the course of this project.}
    \label{fig:timeline}
\end{figure}

We conducted a qualitative study informed by ethnographic methodologies~\cite{o2012ethnographic, harrison2014ethnography}, spanning from March 2024 to March 2025. Our interdisciplinary research team comprises academically trained policy practitioners, scholars of Science and Technology Studies (STS), information scientists, and experts in critical social science methodologies based in Canada. In alignment with established conventions in ethnographic and qualitative research~\cite{higginbottom2013guidance}, we started the study with an open-ended inquiry into the evolving landscape of AI governance in Canada. As fieldwork progressed, our research questions were iteratively refined in response to emerging insights from empirical engagements. In the following sections, we describe how we developed our research objectives, designed our framework, and carried out each step of data collection and analysis in five phases as illustrated in Figure~\ref{fig:timeline}.

\subsection{Mapping Research Objectives and Framework Design}\label{phase1}   
From March to August 2024, we focused on defining research objectives, data collection methods, and analytical frameworks. We aimed to explore how actors in Canada's broader technology policy community perceived AI governance, especially regarding advances in large language models (LLMs). Rather than limiting the scope to AI experts, we chose a wider lens to situate LLMs within broader digital policy debates, including online safety, broadband access, and information integrity. Our protocol underwent rigorous review and was approved from by institutional Research Ethics Board in August 2024. 

Given the lack of a formal definition, we adopted an inclusive understanding of the Canadian technology policy community. This included government officials, tech industry professionals, academic researchers across multiple disciplines, and civil society organizations engaged in digital policy and advocacy. This broad framing informed our participant recruitment strategy in the next phase.

\subsection{Participant recruitment and Data Collection}\label{phase2}  
From September to December 2024, we recruited participants using a mixed sampling strategy: purposive~\cite{ahmad_purposive_2024}, snowball~\cite{noy_sampling_2008}, and convenience sampling~\cite{rezaul_islam_principles_2022}. Participants were selected based on their engagement and experience in Canadian technology policy. Due to funding constraints, participation was not compensated, participation was voluntary. Initial recruitment drew from the authors’ professional networks, built through conferences, research, and advocacy, and expanded via referrals and a public LinkedIn call. Outreach prioritized diversity across government, industry, academia, and civil society, with contact made via email, LinkedIn, or Signal. No strict eligibility criteria were applied beyond relevant experience. Semi-structured interviews, guided by ten core questions, explored participants' experiences with LLMs, AI, views on policy frameworks, critiques of existing governance, and expectations for the future of AI regulation in Canada and globally.

The semi-structured interviews ensured consistency while allowing participants to highlight key issues. Sessions lasted 45–75 minutes via secure video calls and were audio-recorded with consent, then transcribed. Reflexive notes captured additional context. This flexible approach enabled adaptation to emerging themes, ensuring rich qualitative data.

\begin{sidewaystable}
    \centering
\caption{Stakeholder composition and areas of expertise of study participants}
\label{tab:stakeholders}
\begin{tabular}{|l|c|l|}
\hline
\textbf{Stakeholder Group} & \textbf{Number} & \textbf{Type of Expertise} \\
\hline
Civil society organizations & 9 & Human rights, unions \\
\hline
Bureaucrats & 8 & Departments, regulatory agencies \\
\hline
Industry & 6 & Product managers, counsel, data/AI leads \\
\hline
Academia & 12 & Computer science, information science, media studies, law, political science \\
\hline
\end{tabular}
\end{sidewaystable}

\subsection{Preliminary Data Analysis}\label{phase3}  
Between January and March 2025, we analyzed the first 24 interviews, transcribing 26 hours of audio using automated tools followed by manual review to ensure accuracy and anonymized data. The first author also compiled 50 pages of field notes. Using NVivo, an inductive thematic analysis~\cite{mayan_essentials_2023} was conducted with open coding by the first author and weekly team meetings to refine themes. This resulted in 30 initial codes grouped into 12 broader themes. Noting limitations in representational diversity and thematic saturation, the team decided to conduct a second interview round to broaden perspectives within the Canadian technology policy community.

\subsection{Secondary Round of Data Collection}\label{phase4} 
From March to April 2025, a second round of semi-structured interviews was conducted to enhance diversity and analytical depth. Following the review of initial data, the team sought broader sectoral, geographic, and demographic representation within the Canadian technology policy community. Using the same mixed recruitment strategy as Phase 2, 11 additional semi-structured interviews were conducted, informed by emerging themes from the first analysis. Interviews, held virtually, added 12 hours of recordings and 30 pages of notes, strengthening the dataset’s inclusivity.

In total, 35 semi-structured interviews\footnote{The Canadian technology policy community is small and well-connected. To protect anonymity, no demographic details are shared in Table\ref{tab:stakeholders}. Despite diverse perspectives, sharing more identifiers risk participant identification. We appreciate reviewers’ understanding of this ethical choice.} generated approximately 38 hours of audio and over 80 pages of field notes. Participants included eight government-affiliated individuals, twelve academics from diverse disciplines, six private sector professionals from startups to large firms, and nine civil society representatives, including labour unions. This broad range enabled rich, multi-sector qualitative insights into Canadian technology policy.

\subsection{Final Round of Data Analysis}\label{phase5}  
In the final phase (April–May 2025), the team used an inductive approach~\cite{thomas2006general} to analyze anonymized qualitative data. Transcripts and field notes were reviewed iteratively to exclude irrelevant material and highlight key insights. Regular virtual meetings ensured analytical rigour by collectively reviewing discarded content. Relevant excerpts underwent open coding~\cite{strauss1990open} and thematic analysis~\cite{boyatzis1998transforming}, allowing themes to emerge naturally without predefined categories. These were refined into higher-order themes through collaborative rounds, forming the core findings. All data were anonymized, and the study protocol, including recruitment, consent, data handling, and analysis, was approved by a Canadian Research Ethics Board.

\section{Findings}\label{sec5}
Our data reveals a variety of ways in which Canada’s AI policy infrastructure exercises power through control over data, expertise, and algorithms. They also point to the complex and often uncertain ways in which responsible AI norms are being negotiated in response to the opacity and evolving nature of AI technologies. To better articulate the findings, we organize the data into four interrelated themes: (i) expertise, (ii) norms, (iii) accountability, and (iv) imaginaries. Each theme reveals the underlying politics of AI governance and highlights how participants are navigating, negotiating, and at times resisting the tensions embedded within this infrastructure. 

\subsection{Expertise}\label{expertise}
Our discussions with the participants render visible dynamics of power in Canadian AI governance, particularly by outlining which actors are afforded the ability to consult with government and guide policymaking. Our data also reveals the ramifications of these types of expertise-based relationships with the government and their impact on the overall AI governance infrastructure. 

\subsubsection{Sanctioned Expertise Leads to Regulatory Capture}\label{expertise_capture}
Multiple study participants mentioned that a key mechanism of power is the federal government's reliance on external expertise for the iteration of AI governance frameworks and their implementation. External expertise takes the form of contracting consulting firms, select academic researchers, technology company developers, and providers to develop governance products such as national strategies, regulatory frameworks, policies, and guidelines. Participants mentioned that such reliance raised concerns about transparency, accountability, and the erosion of public sector capacity in being able to address issues of technology development. As noted by BFTS scholarship and feminist scholars within STS, prioritizing a specific form of knowledge excludes historically marginalized people and their lived experiences from key moments of decision-making in policy processes~\cite{wajcman_technofeminism_2004, benjamin_captivating_2019, benjamin_race_2019, noble_algorithms_2018, nelson_social_2016}. In the case of AI governance in Canada, expertise was granted to academic researchers with technical knowledge and industry actors who wield a lot of capital power. One of our participants, who works in responsible AI at a technology company, mentioned,

\begin{quote}
    \textit{``I think there's a dependency of outsourcing both the problem solving as well as risk measures, and a lack of leadership in institutions. And all of that outsourcing happens to consulting companies, who effectively, you know, come and tell you what to do."} - Participant 16 
\end{quote}

Several participants cited as an example the auditing framework proposed under Bill C-27, the Artificial Intelligence and Data Act (AIDA)(see Section 15(1) in~\cite{canada_digital_charter}), and argued that its formulation represents how the concentration of recognized expertise within select academic institutions and private sector actors significantly influences the direction of AI policy in Canada. As a result, legislative priorities and problem definitions increasingly reflect the interests and worldviews of these dominant groups, often at the expense of concerns raised by civil society organizations.

As discussed in the background section, AI governance does not solely focus on identifying the technical parameters to minimize discriminatory outputs. Focusing on technical aspects of generative AI tools considers that tech features are the only drivers of systemic discrimination, environmental degradation, insert other problem, copyright infringement, and insert other problem. However, only focusing on those technical issues erases how departments mandates guide their public servants towards adopting technologies that reinforce harms, instead of finding non-technological solutions to the issues they are seeking to address. As a result, these "tech fixes often hide, speed up, and even deepen discrimination" \citep[p.4]{benjamin_race_2019}.

\subsubsection{State Shaping Markets via Funding}\label{expertise_funding}
Interview participants also discussed the federal government's role in funding private sector AI development. For instance, Participant 7, an academic researcher, mentioned that:

\begin{quote}
    \textit{``the private sector has paved the road in how the governance will be and the public sector is essentially subsidizing it."} - Participant 7
\end{quote}

This is exemplified in the consecutive investments for private sector development through the Pan-Canadian AI Strategy launched in 2017, and in the 2024's federal budget \$2.4 billion investments for the creation of the Canadian AI Sovereign Compute Strategy~\cite{1noauthor_securing_2024} as our participants mentioned. While startups are set to receive \$200 million for AI startups to develop technologies that can be adopted in agriculture, healthcare and manufacturing, \$50 million would be set for the creation of the Canadian AI Safety Institute (CAISI), and only \$5.1 million were set for the enforcement of the proposed AIDA, which never became law~\cite{1noauthor_canadian_2024}.

In their responses, participants illustrate how funding is used as an infrastructural tool to shape the political economy of AI in the country, thus reflecting its power to structure the market and commercial priorities. This and similar cases demonstrate how state investment actively shapes the priorities and interactions within the AI ecosystem. From a BFTS standpoint, these funding arrangements can be interpreted as facilitating a regime of wealth accumulation through which marginalized communities' labour, data, and social life, are rendered as sources of value, while simultaneously concentrating power in corporate actors. In turn, this dynamic forecloses opportunities to develop technology in the public interest through participatory and redistributive forms of governance~\cite{hampton_techno-racial_2023}. In sum, Canada’s AI governance infrastructure enacts power in two ways. First, chiefly through expertise by outsourcing decision-making to a narrow circle of sanctioned experts. And second, through resource allocation, by using funding to steer AI development. Combined together, these dynamics of power distribution lead to imbalanced influence and even regulatory capture.

\subsection{Norms}\label{norms}
Our discussions with participants reveal how the rapid development and deployment of LLMs are straining Canada’s existing AI governance norms, particularly in contexts where harms are ambiguous, emergent, or difficult to define. The inherent opacity of many AI systems creates fundamental challenges for establishing shared norms. Participants described the practical difficulty of applying abstract principles such as ``transparency" in the absence of common, operational definitions, underscoring the situated and ongoing work required to make governance actionable.

\subsubsection{Desiring Iteration to Adapt to Technological Changes}\label{norms_iteration}
Participants expressed concerns about the challenge of enforcing existing statutory frameworks, such as those related to copyright, environmental protection, human rights, and labour rights, amid the rapid evolution of large language models. As new capabilities and applications emerge quickly, traditional regulatory mechanisms often lag behind. In response, participants, particularly those working within government departments and technology companies, described their approach to AI governance as inherently iterative. Rather than applying fixed rules, they emphasized the need to continually revise and adapt internal guidelines, policy instruments, and risk frameworks over time. For instance, Participant 8, a government employee, mentions:

\begin{quote}
    \textit{``Our senior executives let us put the first version of our product, it’s going to be helpful, it won’t be perfect, and then we refine it over time."} - Participant 8 
\end{quote}

This adaptive approach reflects a broader logic of governance that aligns with Foucault’s notion of governmentality~\cite{foucault1991governmentality}, wherein power operates through the production of flexible procedures, calculative tools, and institutional routines that manage populations and uncertainty. Governance in this context, as discussed by our participants, is not exercised solely through legal enforcement but through a distributed system of security measures, expert knowledge, and policy experimentation that responds to emerging risks in real time. Iteration, in this sense, becomes a core technique of governing, allowing institutions to demonstrate responsiveness while maintaining control over the framing and management of technological change.

Crucially, this iterative mode of governance also informs how shared norms around responsible AI are established. In the absence of universally agreed-upon principles, norms are not imposed from above but gradually coalesce through repeated rounds of drafting, testing, and adjustment. For instance, Participant 13, a product manager at a technology company, mentions,

\begin{quote}
    \textit{``This like feedback cycle is also something I think that's helpful if there was [...] let's say, someone that I can talk to so that I can bring my version of the, you know, AI policy that I drafted and I discussed with my company [...]  and get some insights from there because a lot of times I feel that we are a little bit in a blind spot. We're all just trying to interpret the law and we tend to go with like, what is the safest."} - Participant 13
\end{quote}

This quote points out that while iteration enables responsiveness and flexibility, it also concentrates normative power in the hands of a limited set of actors, often government insiders and industry stakeholders, who determine the pace, scope, and substance of what responsible AI governance looks like. Rather than resolving uncertainty, iteration manages it, and in doing so, subtly defines the contours of acceptable technological development. Moreover, a BFTS perspective reminds us that iteration emerges as a dual modality. First, iteration can be understood as the mechanisms through which regimes of anti-Blackness and white supremacy continuously adapt and reconstitute themselves, echoing Benjamin's account of race as an ``architecture of power"~\citep[91]{benjamin_race_2019}. Second, iteration can also be understood as a critical praxis of ongoing revision and collective self-reflection through which emancipatory movements continually revise their political commitments and strategies, as articulated in the Combahee River Collective Statement~\cite{combahee2017statement}. As such, while iteration might allow flexibility, growth, and accountability in governance, it may also facilitate the reproduction of harms when used selectively.  

\subsubsection{Negotiating Meaning Amidst Opacity}\label{norms_opacity}
Participants expressed difficulty with conceptualizing what is necessary to govern large language models due to the lack of common, operational definitions, understandings of capabilities, as well as impacts of these technologies. For instance, in reference to struggles with implementing guiding principles, a Participant 23, a government employee working in technology adoption, mentioned that: 

\begin{quote}
    \textit{``It’s another thing to actually implement something effectively, because if we talk about like, say for instance, transparency, what does that mean? Like it could mean something different. To me, it could mean something different to you and things of that nature, right, like when we talk about transparency, people might think, oh, that’s expandability or interoperability or things of that nature. Right. So having a standard set of definitions so that we’re actually pulling from the same language, and we understand what these things mean. I think it’s very critical and very important because unless we’re speaking the same language, the impact is going to be like our ability to kind of implement things in a standardized fashion is going to differ."} - Participant 23 
\end{quote}

In a similar tone, other participants point toward the opacity of AI systems, as also studied in STS~\cite{pasquale_black_2015}, that makes it hard to develop regulatory frameworks. This is because the actors have to negotiate in the absence of key information about the types, capabilities, and impacts of AI systems. A holistic approach would require active consultations, dialogue, and discussions that the participants mentioned were not happening between government, industry, academia, and civil society. Moreover, the difficulty in applying abstract principles of transparency in the absence of shared definitions and understandings of the technology's impacts highlights the lack of efforts in operationalizing them. Building on Frank Pasquale's notion of the \textit{black box}, Benjamin (2019) recasts it through a Black feminist lens as the \textit{anti-Black box}, showing how ostensibly race-neutral technologies, laws, and policies that structure technology development act as tools through which white supremacy operates~\citep[34-35]{benjamin_race_2019}. As such, a BFTS approach to negotiating meaning amid opacity in AI governance interrogates how governance practices themselves sustain and reproduce the technical opacity of AI systems in service of systemic forms of oppression.

\subsubsection{Strategic Illegibility Facilitates Epistemic Injustice}\label{norms_illegibility} 

While participants widely acknowledged the crucial involvement of shared norms in governing LLMs and AI, many also pointed to the persistent lack of transparency across both governmental and corporate institutions as a key obstacle to their formation. Opacity, in this context, is not merely a technical characteristic of AI systems. It is often deliberately maintained and institutionalized. Government agencies and private companies invoke a range of justifications for withholding information, including national security, trade secrecy, and proprietary protections. However, as our participants suggested, such rationales frequently serve to shield practices from public scrutiny and preclude the possibility of democratic oversight. This form of strategic illegibility directly undermines the work required to develop shared governance norms. This strategic illegibility in AI governance mirrors longstanding patterns documented in Black feminist research in Canada, where state violence against Black women, trans, non-binary and gender non-conforming people are routinely rendered invisible due to the government’s selective collection (or deliberate non-collection) of data on policing and enforcement practices~\cite{maynard_policing_2017}. As illustrated in cases documented by legal scholar Gideon Christian, Canadian immigration authorities have attempted to withhold information about using facial recognition technology to revoke refugee status, despite evidence that these biometric surveillance tools disproportionately misidentify Black women~\cite{christian_new_2024, buolamwini_gender_2018}. As Participant 5, a civil society advocate, explained, even legally mandated practices such as privacy impact assessments are often inaccessible:

\begin{quote}
\textit{``Agencies are supposed to engage in privacy impact assessments, but only a summary is shared. Often they’re not shared at all [...] And there’s very [...] little to compel them."} – Participant 5
\end{quote}

Others noted that assessments, when shared, were intentionally obfuscated through legal and technical language or hidden behind claims of confidentiality. Participant 4, an academic researcher, described this as follows:

\begin{quote}
\textit{``Opaque, then they just hide behind a, you know, a wall of obfuscation and proprietary claims of confidentiality."} - Participant 4
\end{quote}

This absence of transparency not only limits critical evaluation of these technologies but effectively blocks the formation of a shared understanding of what ``responsible AI” should entail. This opacity constitutes a form of epistemic injustice~\cite{harding1991whose, russo_epistemic_2025}. In this context, the lack of access to institutional reasoning and decision-making processes effectively excludes civil society, journalists, and researchers from the ongoing negotiation of AI governance norms. Thus, rapid AI advancements are straining Canada’s governance infrastructure by outrunning the creation of shared norms, while the inherent opacity of AI systems creates ambiguity that undermines consistent regulation.

\subsection{Accountability}\label{accountability} 
Across our interviews, participants consistently raised concerns about how Canada’s AI governance infrastructure addresses systemic harms such as scaled bias, discrimination, and environmental degradation linked to the rapid deployment of AI systems. This section focuses on how participants negotiated questions of responsibility and articulated alternative frameworks for accountability.

\subsubsection{Contested Perspectives of AI's Speculative Efficiency}\label{accountability_speculative}
The deployment of AI systems has been done in the name of efficiency, with the expectation that the applications of these tools would increase productivity by facilitating the automation of labour-intensive tasks~\cite{simkute_ironies_2025}. However, there is little evidence of the short, medium, and long-term gains their adoption can bring~\cite{vu_waiting_2024}. This is a tension surfaced during interviews, for instance, when discussing the implementation of these tools in a healthcare context, Participant 9, a civil society advocate mentioned that: 
\begin{quote}
    \textit{``If we're sort of like accelerating the instinct of the person who's the caregiver, the doctor, or the nurse, the person who's like providing the medical care, incentivizes them to not spend the time directly with the person in front of them first and foremost, as the way that they think. It's like you're sort of encouraging a degrading standard of care [...] like that's what you're supposed to be doing every single time. There is an inefficiency in good care."} - Participant 9 
\end{quote}

On the other hand, Participant 22, a government employee, noted that the developments in:
\begin{quote}
    \textit{``natural language processing and large language models have basically brought us closer to that promise, that vision that we were told from the 1950s and 1960s"} - Participant 22 
\end{quote}

These contested views of efficiency, and particularly the potential of these technologies, demonstrate key tensions in how these technologies are viewed by stakeholders interviewed as part of this study. Participants' contested views mirror ongoing academic and societal debates about AI's capabilities and potential. While some researchers anticipate increasingly powerful systems, others critique these claims as over-hyped and unreliable, pointing to an ``AI Snake Oil" phenomena~\cite{narayanan_ai_2024, raji_fallacy_2022}. From a BFTS perspective, these disagreements reveal how assumptions about AI's efficiency are not neutral, but reflect broader power dynamics, including whose expertise is prioritized and whose lives are rendered vulnerable by harmful AI systems~\cite{benjamin_captivating_2019, benjamin_race_2019}. 

\subsubsection{Ambiguous Responsibility Leads to Unclear Channels of Accountability }\label{accountabulity_ambiguity}
Study participants expressed disagreement with identifying which actors and which stages of the AI lifecycle should bear responsibility through legislation and regulation. This reflects struggles over identifying which technical and regulatory components of governance infrastructure play key roles in the outcomes that society experiences when interacting with technologies like AI. For instance, Participant 25, a civil society advocate, linked governance challenges in AI systems to previous challenges experienced with regulating the information ecosystem, given the rise and popularity of social media platforms:

\begin{quote}
    \textit{``go back in time and actually, you know, reform our laws a lot sooner than we have because, you know, we just failed to govern the platform era and the platform may era is what has enabled the AI era"} - Participant 25 
\end{quote}

\begin{quote}
    \textit{``Do you target the developers or do you target the users and we've not had that conversation [...] I personally like the users regulatory approach, because that's where the rubber hits the road. Those are the people who are going to be making decisions that affect people"} - Participant 26
\end{quote}

Participants' remarks echoes Attard-Frost \& Widder (2025) theory on the ethics of AI value chains, which they define as ``as cocreation structures that exist within a network of actors and enable actors to pattern the resource inputs they provide and the resource outputs they receive from AI systems”~\citep[3]{attard-frost_ethics_2025}. They argue that adopting this theory allows governance to develop a regulatory approach that first identifies which actors are to be held accountable, for which actions, and to address challenges beyond the technical components of systems, which widens the scope of governance to respond to environmental and societal impacts of AI. Theories of value chains build on scholarship dedicated to examining and identifying responsibility and accountability throughout the lifecycle of AI systems~\cite{brown_allocating_2023, cobbe_understanding_2023, lee_talkin_2024, widder_dislocated_2023, suchman_located_2002}. In other words, when responsibility is not appropriately identified or accounted for, it opens the door to chains of harm that could have been avoided had actors been properly held accountable. Moreover, adopting a BFTS perspective enables us to trace how AI systems rely on a global supply chain that exploits the Black people in a myriad of ways. From the extraction of minerals in African mines to the underpaid data workers in the Global South for instance~\cite{hampton_techno-racial_2023, noble_future_2016}. As a result, accountability must be traced across every stage of development and use, as the natural resources, materials and labour conditions sustaining AI are inseparable from the techno-racial capitalist systems that exploit Black people globally. 

\subsubsection{Identifying Alternative Accountability Sites}\label{accountability_alternatives}
Study participants recognized the limits of Canada's current regulatory approach and offered alternative methods and practices for sites where accountability could be practised. Notably, Participant 7, an academic researcher, discussed the increasing role of worker-centric collective action as a worthy example of where accountability could be practised. 

\begin{quote}
    \textit{``pay attention to the work that’s coming out of the unions [...] data workers unions, the class action lawsuit on Sama from Kenya [...] the one from the Actors Guild, they won, they won their thing, which kind of shows that they actually managed to ban some stuff [...] that’s really important"} - Participant 7
\end{quote}

This illustrates the dynamics described by scholarship calling to centre the experiences of workers. Particularly, a BFTS perspective highlights how the role of labour organizing has exposed the ways in which companies developing or adopting AI systems in their products enact logics of worker exploitation in the form of psychologically harmful working conditions, wage discrimination and theft, and surveillance~\cite{gebrekidan_content_2024, dubal_new_2021, miceli_data-production_2022, gray_ghost_2019}.

Participant 28, an advocate from a civil society organization, also pointed to union bargaining efforts in Europe as sites where Canada can draw inspiration and learn from. Particularly, they noted that: 
\begin{quote}
    \textit{``The Spanish trade unions involved in this they were even given the right under Spanish law to the algorithms themselves, and when they got the algorithms, they didn't know what to do with it. They couldn't interpret it [...] because that's not where their expertise lay and they had no technical capacity to understand the algorithm from the inside [...] They had to focus on gaining power in the decision making about the application and the purposes to which the algorithms were put."} - Participant 28
\end{quote}

In short, Canada’s AI governance infrastructure currently fails to adequately address AI-specific harms because accountability is diffuse and underdeveloped – ambiguous responsibility and technocratic optimism leave serious issues like bias and environmental impact insufficiently checked.

\subsection{Imaginaries}\label{imaginaries}
While not initially posed as a separate research question, an important emergent theme was how competing visions of the AI future (sociotechnical imaginaries) influence governance. These imaginaries implicitly affect power, norms, and accountability by shaping what futures are prioritized or marginalized. In the following section we consider the different ways participants to the study considered sociotechnical imaginaries about the future of large language models, and particularly how these ways of thinking shape contestations about AI governance infrastructure.

\subsubsection{Imaginaries of Acceleration and Urgency}\label{imaginaries_acceleration} 
Participants to the study raised concerns about the rapidly accelerating advances that AI technologies were making, arguing that such speed made it hard to outline what adequate governance should look like in the country. For instance, Participant 26, a lawyer with experience in government and industry, raised that:

\begin{quote} \textit{``It's a fast moving that you know, if you try to target something right now, in terms of legislation, it just might be outdated."} - Participant 26
\end{quote}

Participants' concerns about the capabilities of these systems point to discourse the techno-deterministic tendencies that conceal how governance is operationalized. Such views frame technological development ``as an external, autonomous force exerting influence on society", and which as a result ``narrows the possibilities for democratic engagement with technology"~\citep[33]{wajcman_technofeminism_2004}. Similarly, these concerns underscore how ``social norms, ideologies, and practices are a constitutive part of technical design"~\citep[41]{benjamin_race_2019}, revealing how governance debates often overlook the social conditions embedded in AI systems themselves.

\begin{quote} \textit{``Right now we're moving into a space where our, you know, our institutions are just not designed to keep up with the change of pace that exists right now. And a lot of that is a product of the fact that we have failed to enforce our existing policy [...] or that the existing enforcement of those policies doesn't have a sufficiently a significant pain for people who are violating those policies" } - Participant 25
\end{quote}

Participant 2, an academic researcher, discussed as well the necessity to change the pace at which AI governance is reacting: 

\begin{quote} \textit{``But what I think that necessitates, which is what we're not taking, is a slower approach to AI regulation and to AI governance. And I think there's a lot of talk about like slow AI, which is great, like we need to slow the pace of AI development. We need to kind of take the foot off the gas a little bit but I think that's also the case in AI regulation and AI policy too."} - Participant 2
\end{quote}

Participants' concerns about the rapid pace of AI development echo Benjamin's (2019) critique of Silicon Valley's ethos, captured in Facebook's original motto ``Move Fast and Break Things," which she reframes with the question, ``What about the people and places broken in the process?"~\citep[13]{benjamin_race_2019}. A slower, more deliberate approach to AI would seek to identify, reflect, address or at least prevent the consequences of technological acceleration, thus highlighting how tech industry's approach to AI development is not trivial, but are ``in fact public policy decisions" due to the AI's profound societal consequences ~\citep[12]{benjamin_race_2019}.

\subsubsection{Imaginaries of Democratic Participation}\label{imaginaries_democracy} 
Participants often raised concerns about the lack of public consultation in Canada's governance efforts, citing its absence as one of they key reasons Bill C-27 and other initiatives received broad public backlash. In response to this government failure, several participants discussed their desire to for participatory democracy forums as a way to ensure governance approaches would be in alignment with the public interest, and receive broad public trust. For instance, Participant 27, an academic researcher, argued that those methods are:

\begin{quote} \textit{``An underrated way of crafting careful and smart policy, and it's so interesting to me that people always say the government takes so long to make policy and we don't want to include participatory democracy or deliberative democracy methods within that policymaking process because it will make the process longer. In my view, I think that's really short-sighted."} - Participant 27
\end{quote}

\begin{quote}\textit{``We need to think about ways to drive engagement and to collect meaningful and informed opinions about technologies [...] definitely worth looking at as models citizens assemblies."} - Participant 1\end{quote}

This echoes calls from critical STS scholars that seek to challenge dominant narratives about technology's inevitability and rapid speed, by advancing alternative visions that emphasize broad public involvement, ethical alignment, and AI serving societal goals~\cite{jasanoff_dreamscapes_2015, haraway_situated_1988, clement_submission_2023}. Moreover, a BFTS perspective on these calls emphasizes the necessity of imagining alternative forms of governance that do more than reforming existing systems. Instead, they must simultaneously seek to dismantle embedded structures of oppression, while actively creating new institutions and practices that embody liberation, echoing W.E.B Du Bois' and Angela Y. Davis' visions of \textit{abolition democracy} and Benjamin's insight that abolition involves both destroying harmful systems and growing new emancipatory ones~\citep[233]{maynard_policing_2017} ~\cite{davis_abolition_2005}~\citep[162]{benjamin_race_2019}. 

In summary, our findings reveal that Canada’s AI governance infrastructure is an active site of negotiation, where expertise, norms, accountability, and imaginaries are continually shaped and contested. Rather than a fixed or coherent system, governance emerges as a dynamic and evolving process, marked by both tensions and efforts to adapt. These insights build on and extend existing scholarship by showing that AI governance must be understood not just through policy instruments or ethical frameworks, but as a sociotechnical infrastructure sustained through care, maintenance, and situated labour, which we discuss in our next section.

\section{Discussion}\label{sec6}
Our findings depict Canada’s AI governance as an active site of negotiation. Far from a coherent system, it is marked by contested expertise, strained norms, and discrepancies in accountability. In light of these challenges, we argue that effectively governing AI’s unique characteristics (opacity, adaptiveness, societal impact) requires a re-conceptualization: viewing ``AI Governance as Infrastructure". As such, in the following section, we apply the critical assessment framework developed in Section~\ref{framework1}. Moreover, we believe that adopting a Black Feminist Technoscience perspective not only allows us to identify the limitations of current governance approaches, but also points to alternative infrastructural designs. As such, we discuss how prioritizing care, maintenance, and adaptive temporalities can foster more equitable and accountable AI futures.

\subsection{Applying the Critical Assessment Framework for AI Governance Infrastructures}\label{applying_assessment}
The following section mobilizes the critical assessment framework we developed in Section~\ref{framework1} to make sense of how participants experience and navigate AI governance in practice. Rather than treating these dimensions as abstract categories, we show how they emerge relationally through interviewees’ descriptions of power, norms, accountability, and imagined futures. More specifically, we demonstrate how:
\begin{enumerate}
    \item \textit{Power Configuration}: Corresponds to the politics of expertise we found in Section~\ref{expertise}, identifying how power is distributed and challenged within the governance infrastructure.  
    \item \textit{Governance Pressures}: Assesses the strain of rapid and opaque AI development in institutions discussed as discussed in Section~\ref{norms}. 
    \item \textit{Accountability Gaps}: Captures the failings we observed in addressing harms through the formal structures of accountability in Section~\ref{accountability}.
    \item \textit{Contested Futures}: Identifies how actors in the infrastructure envision AI futures, as articulated in Section~\ref{imaginaries}.
\end{enumerate}

\subsubsection{Power Configuration} 
AI governance infrastructure includes not only laws and policies but also the actors shaping them in government, industry and civil society. Broadening the definition of expertise helps shift governance away from a political economy that prioritizes technology adoption towards the public interest. As noted in the findings, AI’s political economy focuses on capital investments in computing and labor~\cite{couldry_costs_2020}, particularly those from actors that are financially invested in the growth and adoption of AI. Excluding non-industry voices risks reinforcing governance that favours capital over human rights, labour, and the environment~\cite{brandusescu_artificial_2021, bannerman_tech_2024, levesque_smoke_2025}. To be developed, responsibly adopted, and maintained, AI systems require significant amounts of concerted labour from software engineers, product managers, compliance officer, policymakers, academic researchers across disciplines, and custodial staff who maintain physical infrastructure such as offices and data centres for instance. As such, infrastructure of AI governance requires the mobilization of regulatory resources that reflect the broad range actors involved.

\subsubsection{Governance Pressures} 
In Section~\ref{norms}, we illustrate how participant responses suggest that actors in Canada's technology policy community engage in constant and difficult negotiations to identify the capabilities of AI systems and corresponding policy frameworks. We find that the opacity of AI systems adds an additional dimension of difficulty, as it makes it hard for actors part of the infrastructure to develop shared understanding of the technology's capabilities. Moreover, the opacity of AI systems is used as an excuse for government's to engage in a strategic use of illegibility. Consequently, the opacity of AI systems is replicated by governance infrastructure, allowing AI opacity to not just be a technical feature, but also a political one. 

\subsubsection{Accountability Gaps}
In Section~\ref{accountability}, the lack of accountability \cite{noble_algorithms_2018, benjamin_race_2019} and the replication of the tool's technical opacity to its governance infrastructure \cite{pasquale_black_2015}. Participants to the study discussed how formal channels of communication with government were inadequate due to their exclusionary nature, thus marginalizing discussions about developing alternative methods for accountability to address governance failures. When reflecting about alternative methods of accountability, participants often mentioned the potential of labour organizing, often citing examples from workers in data centres, technology companies, and in the creative industry as worthy examples to follow. Similarly, Black Feminist theory scholarship recognizes the potential for resistance, justice-making, accountability and sustainability through labour organizing \cite{williams_exploited_2022, davis_women_1983, combahee2017statement}.

\subsubsection{Contested Futures} 
This study shows how competing sociotechnical imaginaries \cite{jasanoff_dreamscapes_2015, jasanoff_containing_2009} can undermine public participation and marginalize dissent in AI policy. Participants highlighted the promise of participatory democracy—through citizen assemblies, public forums, and commissions—as a way to engage the public in deciding whether and how AI should be used, and to hold institutions accountable. A key example is the Canadian Commission on Democratic Expression (CCDE), launched in 2020 to examine digital technologies’ impact on democracy~\cite{1centre_for_media_technology_and_democracy_bill_nodate}. Over three years, it held 42 citizen assemblies that produced policy recommendations, prompting government consultations and the development of Bill C-63. Though the bill didn’t pass, the process shows that participatory democracy is not only viable but can foster more responsive policymaking.

\subsection{Towards Sustainable AI Governance Infrastructure}
Having identified the key problem areas (power, governance pressures, accountability, and contested futures) in Section~\ref{applying_assessment}, we now turn to how a Black Feminist Technoscience lens can inform alternative governance approaches. A critical infrastructural analysis, informed by BFTS allows ys to not only identify and understand the problems with AI governance in Canada, but also allows us to formulate alternative approach which center equity, accountability, justice, and sustainability. In particular, we discuss how embedding care, exercising refusal, prioritizing maintenance, and embracing adaptive temporality in AI policy. A critical infrastructural analysis, informed by Black Feminist Technoscience allows us to not only identify and understand the problems with AI governance in Canada, but also allows us to formulate alternative approaches which center equity, accountability, justice, and sustainability. To achieve this requires shifting governance goals away from the promotion of innovation, towards practices of care, maintenance, repair, and acknowledging adaptive temporality. 

\subsubsection{Embedding Care}
Care can be broadly defined as the political and relational practice of attending to the needs of people, creatures, nature, objects, phenomena, or ideas for instance~\cite{Stephen2026Care, martin_politics_2015, puig_de_la_bellacasa_matters_2017}. Care is also fraught, as scholars acknowledge it has been used to categorize, discipline, and regulate people~\cite{martin_politics_2015, maynard_policing_2017, Jones2022AbolitionistIntimacies, Roberts2022TornApart, Stephen2026Care, hooks_all_2018}. Consequently, an AI governance infrastructure grounded in the ethics of care, requires moving beyond abstract principles of fairness, equity, diversity and inclusion that foreground current policy and scholarly debates~\cite{birhane_algorithmic_2021}. While STS scholars urge researchers to consider "how care operates in sociotechnical contexts, but also to the roles that we play in our studies of technoscience and our accountabilities to the worlds that we construct"~\citep[629]{martin_politics_2015}, we extend this call to actors part of the Canadian AI Governance infrastructure. Particularly, we want to push policymakers to consider their positionality, as well as their power vis-à-vis other actors in the infrastructure. 

Care in AI governance is not only enacted through attentive design or consultation, but also through deliberate acts of refusal~\cite{Stephen2026Care}. Across a range of research disciplines, Black, Indigenous and racialized scholars all over the world have worked towards theorizing refusal as an important mode of resistance~\cite{adelman_technoskepticism_2025, benjamin_informed_2016, pena_gangadharan_digital_2021}. This idea of refusal – saying no to technologies that undermine care or equity – challenges the inevitability narrative around AI and creates space for more considered, community-oriented decision-making. Applied to AI governance, refusal entails not only safeguarding individuals’ ability to decline technological use in domains such as employment, education, and healthcare, but also ensuring that there are independent mechanisms available within government institutions to proactively abstain from developing, deploying, or endorsing technologies that violate human rights charters, undermine climate commitments, or conflict with obligations under the United Nations Declaration on the Rights of Indigenous Peoples' for instance. Left to the discretionary goodwill of policymakers, practices of refusal remain fragile and reversible; institutionalizing refusal through independent infrastructural mechanisms capable of resisting political volatility and the recurrent hype cycles that characterize AI development~\cite{rudschies_long_2025, moon_promises_2026, mcconvey_risk_2026}.

\subsubsection{Prioritizing Maintenance and Repair}
STS scholars emphasize that infrastructures require ongoing maintenance and repair to remain functional and equitable~\cite{gillespie_rethinking_2014, murphy_sick_2006}. In AI, this means enacting legislation that mandates continuous monitoring, evaluation, and repair of the governance infrastructure. It involves adopting meaningful information disclosure—beyond audits or privacy impact assessments—to include financial reporting, public consultation strategies, and other transparency measures. Robust regulatory oversight is essential to prevent transparency theatre and ensure that policies are regularly revised to address evolving harms and contexts. Moreover, from a BFTS perspective, maintenance does not just call for large-scale federal policy changes. A BFTS perspective on maintenance brings attention to how localized, community governance initiatives, are also crucial components of the AI governance infrastructure. Resisting AI adoption in mutual-aid organizing or establishing privacy-preserving alternatives to cloud data storage for a neighbourhood's health clinic for instance, are all important, yet overlooked and underappreciated sites of AI governance where maintenance enables the cultivation of alternative infrastructural designs. 

Crucially, sustained governance repair is needed to build public trust. Civil society organizations have criticized Canada's proposed \textit{AI and Data Act} for lacking public consultation and transparency~\cite{1csilc_key_2024, tessono_ai_2022}, while the legislative process around Bill C-27 was marked by delays and dysfunction~\cite{benson_devils_2023}. In the same way that new technologies can reinforce and interlock systemic forms of oppression~\cite{benjamin_captivating_2019}, it is important to identify how AI governance perpetuates those patterns as well in attempts to foster public trust.

\subsubsection{Acknowledging Temporality and Adaptability}
Companies are developing new capabilities for various AI systems everyday, researchers are reporting new breakthroughs through the adoption of these technologies. However, a key problem that surfaced during our interviews with experts across academic disciplines, civil society advocacy organizations, government institutions, and industry, is the fear that that policy infrastructure wouldn't be able to withstand the pace at which these technologies are entering markets. This framing indicates that innovation follows a singular trajectory, one marked my rapid and disruptive technological development and adoption and that governance must try to chase it in a similar fashion. Yet, BFTS perspectives on Black temporality challenge this assumption. For instance, Sherman (2023), drawing on Ibrahim \& Ahad (2022), argues that technology innovation presumes "a linearity of time that is only forwardly progressing", and that by contesting this "single story of design", alternative futures can be imagined~\citep[982]{sherman_black_2023}~\cite{ibrahim_introduction_2022}. As such, a BFTS perspective reframes the problem: rather than treating technological innovation as the pace that governance must chase, it centers the cultivation of governance capacity, shifting from singular policy fixes towards ongoing processes of governance. BFTS facilitates building infrastructural designs which include, but are not limited to periodic review, co-governance, sunset clauses, as a way to ensure governance remains an ongoing process, collective process rather than one that only reacts to disruption.

\subsection{Heuristic Policy Guidance for Sustainable AI Governance Infrastructure}
To address issues of the continuous contestations, opacity in policymaking, and accountability gaps, and to facilitate adaptability, care, maintenance, repair, and refusal, we developed a series of policy recommendations. The following five recommendations are grounded in our findings and the above critical framework, aiming to address the power imbalances, opacity, and accountability gaps identified. They also incorporate BFTS values like participatory democracy, care, and sustainability.

\begin{enumerate}
    \item \textbf{Establish independent regulatory oversight} An independent body would address Canada's overreliance on the private sector and lack of regulatory independence in AI governance. It would have broad powers to investigate, audit, fine, recall, prosecute, ban, and standardize AI systems, ensuring safety before market entry and enabling accessible accountability when harms occur, thus addressing the risk of regulatory capture as discussed in Section~\ref{expertise}.
    \item \textbf{Enact mandatory public reporting to counter opacity} Requiring federal departments and agencies to publicly disclose in a timely manner AI-related procurement details, audit outcomes, and impact assessments. This aims to counter the replication of AI systems’ opacity within governance structures, which limits public accountability. Consequently, effective and robust mandatory reporting measures would address the concerns of participants raised in Section~\ref{norms}.
    \item \textbf{Empower worker-led initiatives as alternative sites of accountability} Empowering workers to define and determine how technologies affect their jobs. Practically, this could involve amending the labour laws to establish technology adoption councils, similar to Germany’s worker councils~\cite{cukier_workers_2023, worksconstitution_germany}. This recommendation facilitates proposals which participants raised in Section~\ref{accountability}. 
    \item \textbf{Embed participatory democracy to facilitate public interest technology} This includes establishing mandatory consultative offices within government bodies to ensure consistent public input beyond ministerial discretion. Effective participation requires ongoing evaluation, inclusive engagement (both in-person and remote), fair compensation, and timely publication of consultation outcomes. This recommendation addresses calls for better democratic engagement participants expressed in Section~\ref{imaginaries}.
    \item \textbf{Legislate co-governance for regulatory agility} Digital regulator forums are a way to enable co-governance and cross-sectoral collaboration on emerging technologies like AI~\cite{attard-frost_ai_2024, renieris_best_2023, cramer_digital_2024}. Moving toward co-governance beyond information sharing would empower regulators to address complex digital issues more effectively. A legislated forum would grant regulators the flexibility, powers, and structure to address intersecting digital issues more proactively. This recommendation operationalizes better governance practices and speaks to the concerns raised by participants during interviews. 
\end{enumerate}

\section{Limitations and Future Research}\label{sec7} 
Interviews were conducted from September 2024 to April 2025, a period marked by rapid shifts in AI governance and global trade tensions, including a trade war between Canada and the U.S.~\cite{lord_us_2025}. Participant views may have evolved during this time, but we did not conduct follow-up interviews. Additionally, the sample does not represent the full spectrum of Canada’s tech policy community, and the study does not aim to offer an exhaustive or representative account of AI governance perspectives. This is exemplified by the smaller number of industry actors interviewed, in contrast to other groups such as academia and civil society. Future research should adopt a relational lens to examine how actors interact and what incentives shape their behaviour, as explored by~\cite{ahmed_field-building_2024} in their study of AI safety's epistemic culture. All interviews were conducted in English for practical reasons, including transcription ease and limited team fluency in French. Future studies should examine AI governance discourse in both official languages to identify linguistic trends or disparities, particularly since English-language research is not always available in French. Beyond Canada, future studies could apply our framework to evaluate how AI governance infrastructure is configured in other jurisdictions or even globally.

\section{Conclusion}\label{sec8}
This paper presents a critical framework for viewing AI regulatory strategies as ``Governance as Infrastructure." Drawing from Infrastructure Studies, Science and Technology Studies, and Black Feminist Technoscience, we analyzed Canadian AI governance through interviews with the technology policy community. Our analysis showed how AI governance enforces power through politics of expertise, how actors manage AI’s opacity often via strategic ignorance, and how accountability mechanisms fall short in addressing AI-specific harms. We offer recommendations that point to alternative governance models centering care, maintenance, participatory democracy, and co-governance to create fairer, more accountable AI futures. Together, these insights demonstrate the value of viewing AI governance as a sociotechnical infrastructure. This lens, especially when combined with a Black Feminist Technoscience perspective, uncovers hidden dynamics of power and exclusion that typical policy analyses overlook. By centering care, refusal, maintenance, and adaptive temporality, our approach points toward a fundamentally different paradigm for AI governance: one that other jurisdictions might also consider as they grapple with similar challenges.

\bibliography{references}

\end{document}